\documentstyle[preprint,eqsecnum,aps,12pt]{revtex}
\def\vec#1{\stackrel{\rightarrow }{#1}}
\author {T. Garel}
\address{ CEA, Service de
Physique Th\'eorique\\ CE-Saclay\\ 91191 Gif-sur-Yvette, Cedex\\ France}
\author{Mehran Kardar\footnote
{Permanent address: Department of Physics, Massachusetts Institute of
Technology,
Cambridge, Massachusetts 02139, USA}
and H. Orland\footnote
{Permanent address: Service de Physique Th\'eorique, CE-Saclay and
Groupe de Physique Statistique, Universit\'e de
Cergy-Pontoise, 95806 Cergy-Pontoise Cedex, France}
}

\address{ Institute for Theoretical Physics\\ University of California\\
Santa Barbara\\ California 93106, USA}

\title{ADSORPTION OF POLYMERS ON A FLUCTUATING SURFACE}

\begin{document}

\maketitle
\begin{abstract}
We study the adsorption of polymer chains on a fluctuating surface.
Physical examples are provided by polymer adsorption at the rough
interface between two non-miscible liquids, or on a membrane. In a
mean-field approach, we find that the self--avoiding chains undergo an
adsorption
transition, accompanied by a stiffening of the fluctuating surface.
In particular, adsorption of polymers on a membrane induces a
surface tension and leads to a strong suppression of roughness.
\end{abstract}

\bibliographystyle{unsrt}

\noindent \mbox{PACS: 61.41, 68.10, 61.25.H} \newpage
The adsorption of polymers on a surface is of both technological and
biological importance. The coating of a {\em rigid} wall by polymers has been
studied extensively\cite{PGG79}. The adsorption of polymers (e.g. proteins)
on a cell membrane is further complicated by the fluctuations of the
adsorbing surface. This problem bears some resemblance to the wetting
of a substrate\cite{Dietrich},  which is easier to study due to the
absence of self--avoiding constraints for the wetting interface\cite{Binder}.
The latter problem has been studied more recently in the presence of
(quenched or annealed) roughness of the adsorbing
surface\cite{ForOrlSch.1,AndJoa.1,LiKardar}.

In this letter, we study the adsorption of a single polymer, or a dilute
solution of polymer chains, on a {\em soft} (fluctuating) surface. A
related case has been studied by de Gennes \cite{PGG90}.
Examples of soft surfaces are provided by the boundary of a liquid, or
the interface between two non-miscible fluids below their demixing point.
Fluctuations of the surface in these examples are controlled
by the surface tension. Other examples are fluid (or polymerized) membranes,
where fluctuations are governed by the resistance to bending (or anomalous
rigidity).
An ideal polymer chain is always adsorbed to an attracting surface, hard
or soft. Using a variational approach, we show that there is an unbinding
transition for {\em self--avoiding} polymers. The critical temperature $T_c$,
is inversely proportional to the excluded volume parameter $v$, and to the
surface concentration of monomers $c$. The adsorption of polymers
at temperatures below $T_c$ is accompanied by a stiffening of the surface.
This effect is most pronounced for fluid membranes, which acquire an
effective surface tension from the adsorbed polymer layer.

Consider a fluctuating $D$ dimensional ``surface" embedded in $d=D+1$
dimensional space. Within the SOS (solid-on-solid) approximation,
configurations of the surface are parametrized by its height $z({\bf x})$,
where ${\bf x}$ is a $D$ dimensional vector spanning the $z=0$ plane.
The surface has linear dimension $L$ and an area that scales as $L^D$.
We assume that it is in contact with a dilute solution of ${\cal N}$
chains of $N$ monomers each, such that the total number of monomers
$N_m={\cal N}N$, also
scales as $L^D$. The conformations of the polymers are indicated by
$d$ dimensional vectors $\{\vec{r_i}(s)\}$, where
$s$ denotes the curvilinear abscissa of a monomer of Kuhn length $a$.
Using a coarse grained Hamiltonian for the long wavelength
fluctuations, the partition function is
\begin{eqnarray}\label{partition}
Z=\int {\cal D}z({\bf x})\ \prod_{i=1}^{{\cal N}}{\cal D}
\vec{r_i}(s)&&\exp \left[ -\frac K2\int^Ld^D{\bf x}
\left( \nabla ^n z({\bf x})\right) ^2-\ \frac d{2a^2}\sum_{i=1}^{{\cal N}}
\int_0^Nds\ \left( \frac{d\vec {r_i}}{ds}\right)^2\right]\nonumber\\
\times &&\exp \left[-\frac v2\sum_{i,j=1}^{{\cal N}}\int_0^Nds\int_0^N
ds^{\prime}\ \delta(\vec {r_i}(s)-\vec
{r_j}(s^{\prime }))\right]\nonumber\\
\times &&\exp \left[ \ \beta \ \int^Ld^D{\bf x}
\sum_{i=1}^{{\cal N}}\int_0^Nds\ \ \delta (\vec
{\rho}(\vec{\bf x})-\vec{r_i}(s))\right].
\end{eqnarray}

The first two terms in the above expression denote respectively the elastic
energy costs of distorting the surface and the polymers.
The exponent $n$ is equal to $1$ for interfaces (surface tension dominated),
and $2$ for fluid membranes (bending energy). The third term represents the
repulsion between monomers in a good solvent ($v>0$ is the excluded volume
parameter). The final term mimics a short range attraction between the
monomers and the surface (with $\beta=1/T$). The vector
$\vec {\rho }({\bf x})=({\bf x},\ z({\bf x}))$ represents
a generic point on the surface, and we have assumed that a monomer at
$\vec {r}_i(s)$ interacts with the  point
$\vec {\rho }({\bf x})$ on the surface via a short range
($\delta$ function) attraction. (The bound $L$ on the ${\bf x}$ integrals
represents the upper cut-off due to finite surface size.)

By performing standard Gaussian transforms, the interaction terms in
eq.(\ref{partition}) can be rewritten, yielding
\begin{eqnarray}\label{Gauss-trans}
Z=&&\int{\cal D}\Phi(\vec{r})\int{\cal D}(\Psi^{\dagger}
(\vec{r}),\Psi(\vec {r}))\
\exp\left[ -\int d^dr\ \left(\frac 12\Phi ^2(\vec {r})+
\Psi^{\dagger}(\vec{r})\Psi(\vec{r})
\right)\right] \nonumber\\
&&\int{\cal D}z({\bf x})\ \exp \left[-\int^Ld^D{\bf x}\ \left(
\frac K2\left(\nabla^n z({\bf x})\right)^2-\sqrt{\beta}\ \Psi({\bf x},\ z({\bf
x}))
\right)\right]\nonumber\\
&&\int\prod_{i=1}^{N}{\cal D}\vec{r_i}(s)
\exp \left[-\sum_{i=1}^{{\cal N}}\int_0^Nds\  \left(
\frac d{2a^2}\left( \frac{d\vec {r_i}}{ds}\right)^2
+i\sqrt{v}\Phi (\vec {r_i}(s))\
-\sqrt{\beta}\Psi^{\dagger}(\vec{r_i}(s))
\right)\right].
\end{eqnarray}
The direct interactions of the polymer coordinates have now been removed
and the final integration describes the motion of polymers in an
effective potential \hbox{$i\sqrt{v}\Phi(\vec r)
-\sqrt{\beta}\Psi^{\dagger}(\vec r)$}.
The polymeric contribution is thus equivalent to the path integral for the
evolution of
${\cal N}$ quantum particles moving in this effective potential\cite{Edwards}.
This analogy indicates that, in the limit of long
chains $N\rightarrow\infty$, properties of the path integral are dominated by
the ground state of the particle in the potential\cite{Wiegel}.
In a variational treatment of the corresponding Schr\"odinger equation, the
ground state wave-function and energy are obtained from
\begin{equation}\label{varE}
E_0=\min\left\{  \int d^dr\ \left[ \frac{a^2}{2d}
\left( \nabla \varphi (\vec {r})\right)^2+
\left(i\sqrt{v}\Phi(\vec r)-\sqrt{\beta}\Psi^\dagger(\vec r)
\right)\varphi(\vec {r})^2-E\varphi(\vec {r})^2
\right]\right\}_\varphi,
\end{equation}
where, in the final term, the Lagrange multiplier $E$ is used to enforce the
normalization condition on the wave-function $\varphi(\vec r)$.

After substituting $\exp\left\{-N_m E_0[\Phi,\Psi^\dagger,z]\right\}$ for the
polymeric
integrals in eq.(\ref{Gauss-trans}), the functional integrals over $\Phi$,
and $(\Psi,\Psi^\dagger)$ can be performed to yield
\begin{equation}\label{ground-state}
Z=\int{\cal D}z({\bf x})\ \exp \left[ -\int^Ld^D{\bf x}\left(
\frac K2\left(\nabla^n z({\bf x})\right)^2+N_mU\left[z({\bf x})\right]
\right)\right] ,
\end{equation}
where $N_m=N{\cal N}$ is the total number of monomers, and
\begin{equation}\label{U}
U\left[z({\bf x})\right]=\min\left\{\int d^dr\ \left( \frac{a^2}{2d}
\left( \nabla \varphi (\vec {r})\right)^2+
\frac{N_mv}2\varphi^4(\vec {r})
-E\varphi^2(\vec {r})\right)
-\beta \int d^D{\bf x}\ \varphi^2({\bf x},\ z({\bf x}))\right\}_\varphi .
\end{equation}
For a uniform surface $z({\bf x})=z_0$, translational invariance requires that
the
ground state wave-function have the form $\varphi (\vec {x},z)\sim
\phi(z-z_0)$.
We make the assumption that in the presence of a small surface distortion
$z({\bf x})$, the minimum of eq.(\ref{U}) is only weakly modified and is
self--affinely
distorted to
\begin{equation}\label{trans-inv}
\varphi ({\bf x}, z)={\frac 1{L^{D/2}}}\ \phi (z-z({\bf x})) .
\end{equation}
With the added scaling factor in eq.(\ref{trans-inv}), the new wave function
must be normalized such that $\int dz \phi(z)^2=1$.

With the above ansatz, the only dependence of $U[z({\bf x})]$ comes from the
$(\nabla\varphi)^2$ term in eq.(\ref{U}), and leads to
\begin{equation}\label{effZ}
Z\propto\int{\cal D}z({\bf x})\ \exp \left[ -\int^Ld^D{\bf x}\left(
\frac K2\left(\nabla^n z({\bf x})\right)^2+\frac{ca^2}{2d}\left(\nabla z({\bf
x})\right)^2
\int dz\left( \frac{\partial \phi }{\partial z}\right) ^2\right) \right] .
\end{equation}
We can now perform the Gaussian integrals over $z({\bf x})$ to obtain
\begin{eqnarray}\label{free-energy}
\frac{\beta F}{L^D}=\min\Bigg\{\frac 12\int \frac{d^D{\bf k}}{(2\pi )^D}\ln \
\left( K{\bf k}^{2n}+
\ \frac{ca^2}d  {\bf k}^2\int dz\left( \frac{\partial \phi }{\partial z}\right)
^2\right) +
\frac{ca^2}{2d}\int dz\left( \frac{\partial\phi }{\partial z}\right) ^2
\nonumber\\
-c\beta \int dz\ \delta (z)\ \phi^2(z)+\frac{vc^2}2\int dz\ \phi ^4(z)-cE\int
dz\ \phi^2(z)
\Bigg\}_\phi ,
\end{eqnarray}
where $c=N_m/L^D$ is the monomer concentration per unit area of the interface.
The minimization with respect to $\phi$ yields the differential equation
\begin{equation}\label{equa-diff}
-\frac{a^2}{2d}\left[1+\int \frac{d^D{\bf k}}{(2\pi )^D}
\frac{{\bf k}^2}{\left( K{\bf k}^{2n}+
\frac{ca^2}d{\bf k}^2\int dz\left( \frac{\partial \phi }{\partial
z}\right)^2\right) }\right]
\phi ^{^{\prime \prime}}(z)-\beta \delta (z)\phi (z)+vc\phi^3(z)=E\phi (z) ,
\end{equation}
which is similar to the equation describing polymer adsorption to an attractive
potential
at $z=0$, with a renormalized Kuhn length\cite{PGG79}.

The non-linear self-consistent equation (\ref{equa-diff}) is still too
complicated to be solved
exactly. Instead we use a restricted class of wave functions, namely normalized
Gaussians
\begin{equation}\label{gaussian}
\phi ^2(z)=\frac{1}{\sqrt{2\pi w^2}}\exp\left(- \frac{z^2}{2w^2}\right),
\end{equation}
with the width $w$ as the only variational parameter,  to minimize
the free energy in eq.(\ref{free-energy}). (This approximation works quite well
in
other adsorption problems.) The calculations are now straightforward, and we
obtain
\begin{equation}\label{free-gaussian}
\frac {\beta F}{L^D}=\min\left\{\frac 12\int \frac{d^D{\bf k}}{(2\pi )^D}\ln \
\left( K{\bf k}^{2n}+\ \frac{ca^2}{4dw^2}{\bf k}^2\right) +\frac{ca^2}{8dw^2}-
\frac{c\beta }{\sqrt{2\pi }w}+ \frac{vc^2}{4\sqrt{\pi }w} \right\}_w.
\end{equation}
The extremal values of $w$ satisfy
\begin{equation}\label{eq-w}
\frac{a_R^2}{4d\ w^3}+\frac{cv}{4\sqrt{\pi }w^2}=\frac \beta {\sqrt{2\pi }w^2}
,
\end{equation}
with
\begin{equation}\label{a-Rw}
a_R^2=\left( 1+\int \frac{d^D{\bf k}}{(2\pi )^D}\ \frac{{\bf k}^2}{K{\bf
k}^{2n}+
\ \frac{ca^2}{4dw^2}{\bf k}^2}\right) a^2 .
\end{equation}
Note that, within this approximation, the height-height correlation function
of the surface is
\begin{equation}\label{corr-func}
\left\langle \left|z({\bf k})\right|^2\right\rangle =
\frac 1{K{\bf k}^{2n}+\ \frac{ca^2}{4dw^2}{\bf k}^2},
\end{equation}
leading to a surface width
\begin{equation}\label{width}
\gamma =\int \frac{d^D{\bf k}}{(2\pi )^D}\ \frac 1{K{\bf k}^{2n}+\
\frac{ca^2}{4dw^2}{\bf k}^2}.
\end{equation}

In the following, we shall focus on the surface in three dimensional space
($d=3$ and
$D=2$). Independent of $n$, the system undergoes a second order phase
transition at
\begin{equation}\label{temp-crit}
T_c=\frac{2\sqrt{2}}{cv}.
\end{equation}
At high temperatures, the polymer is not adsorbed and eq.(\ref{eq-w}) is
minimized for
\begin{equation}\label{high-T}
w \to+\infty .
\end{equation}
The fluctuating surface is rough, with a width which diverges with its size $L$
as
\begin{equation}\label{gamma1}
\gamma =\frac 1{2\pi K}\ln (\Lambda L)\qquad\text{for }\, n=1,
\end{equation}
where $\Lambda$ is a short distance cut--off, and
\begin{equation}\label{gamma2}
\gamma =\frac 1{4\pi K}L^2\qquad\text{for }\, n=2.
\end{equation}
Below the critical temperature, the polymer is adsorbed on the interface
and $w$ is finite. Note that the critical temperature decreases when the
polymer concentration or excluded volume increases. In particular, Brownian
chains are {\it always} adsorbed to the surface.

The critical behavior is easily obtained from eqs.(\ref{eq-w}) and
(\ref{a-Rw}).
Defining the reduced temperature $t=T_c-T,$ we find:
\par\noindent
{\bf (i)} For an elastic interface ($n=1$),
\begin{equation}\label{crit-behav}
w  \sim \frac 1{v^2c^2}t^{-1}\qquad{\rm and}\qquad \gamma =\frac 1{2\pi
K_R}\ln (\Lambda L),
\end{equation}
where the stiffness $K_R$ behaves as
\begin{equation}\label{crit-stiff}
K_R\sim K+c\ t^2 ,
\end{equation}
close to the transition.
On the other hand, at low temperatures the width of the polymer layer goes
to zero as $w\sim T$ and  $K_R$ diverges as $cT^{-2}$ .
\par\noindent
{\bf (ii)}  For the fluid membrane ($n=2$),
\begin{equation}\label{crit-behav2}
w  \sim -\frac 1{v^2c^2}t^{-1}\ln \ t \qquad{\rm and}\qquad
\gamma \sim \frac{(\ln \ t)^2}{c^5v^4t^2}\ \ln \ L .
\end{equation}
Although there is no surface tension in the free system, the adsorption of
polymers produces a stiffness for the membrane. The critical behavior of the
width of the adsorbed layer  $w$, is larger by a logarithmic factor than for an
elastic
membrane. The overall fluctuations of the membrane $\gamma $, are strongly
reduced
due to the induced stiffness.

To summarize, we have studied (in a mean-field like approximation) the
adsorption of (self--avoiding) polymer chains on a fluctuating surface.
At low temperatures the polymers are adsorbed in a thin layer to the surface,
while at high temperatures they desorb and move freely into the solution. The
adsorption of polymers stiffens the surface and reduces the extent of
transverse
fluctuations. Adsorption of polymers thus provides a useful mechanism for
controlling the fluctuations of membranes, as well as the degree of swelling in
lamellar phases. Presumably more rigid polymers are even more efficient in
controlling fluctuations; the extreme limit of polyelectrolytes being quite
interesting.
Other extensions of this work could be the study of the adsorption  of
copolymers (block or statistical), or different types of polymers.

We would like to thank D. Andelman and
E. Guitter for useful discussions and comments.
This research was supported in part by the National Science Foundation under
Grant No. PHY89-04035. MK acknowledges support from the NSF through grants
DMR-93-03667 and PYI/DMR-89-58061.

\bibliography{myref}

\begin{thebibliography}{99}
\bibitem{PGG79}  P.G. de~Gennes, \newblock {\em Scaling concepts in polymer
physics} \newblock (Cornell University Press, Ithaca, 1979).

\bibitem{Dietrich}  S.~Dietrich, \newblock in  {\em Phase Transitions and
Critical Phenomena}, volume~12, C.~Domb and J.L. Lebowitz, editors (Academic
Press, London, 1988).

\bibitem{Binder}  K.~Binder, \newblock in {\em Phase Transitions and Critical
Phenomena},
volume~8, C.~Domb and J.L. Lebowitz, editors (Academic Press, London, 1988).

\bibitem{ForOrlSch.1}  G.~Forgacs, H.~Orland, and M.~Schick, \newblock
{\em Phys. Rev. B} {\bf 32}, 4683 (1985).

\bibitem{AndJoa.1}  D.~Andelman and J.-F. Joanny,  and M.O. Robbins,
\newblock {\em Europhys. Lett.} {\bf 7}, 731 (1988).

\bibitem{LiKardar} H. Li and M. Kardar, \newblock {\em Phys. Rev. B} {\bf 42},
6546 (1990).

\bibitem{PGG90} P.G. de Gennes, \newblock {J. Phys. Chem.} {\bf 94},
8407 (1990).



\bibitem{Edwards}  S.F. Edwards, \newblock in  {\em Path Integrals}, page 285,
G.J. Papadopoulos and J.T. Devreese, editors (Plenum, New York, 1978).

\bibitem{Wiegel}  F.~Wiegel, \newblock {\em Phys. Rep.} {\bf 16}, 59 (1975).


\end{thebibliography}

\end{document}